\documentclass[12pt]{article}

\begin{document}
\begin{center}
{\bf TRACE ANOMALY AND QUANTIZATION OF MAXWELL'S THEORY ON
NON-COMMUTATIVE SPACES}\\
\vspace{5mm}
 S.I. Kruglov \\
\vspace{5mm}
\textit{International Educational Centre, 2727 Steeles Ave. W, \# 202, \\
Toronto, Ontario, Canada M3J 3G9}
\end{center}

\begin{abstract}
The canonical and symmetrical energy-momentum tensors and their
non-zero traces in Maxwell's theory on non-commutative spaces have
been found. Dirac's quantization of the theory under consideration
has been performed. I have found the extended Hamiltonian and
equations of motion in the general gauge covariant form.
\end{abstract}

\section{ Introduction}

Non-commutative (NC) gauge theories are of interest now because
they appear in the superstring theory \cite{Seiberg1999}. In the
presence of the external background magnetic field, NC coordinates
emerge naturally \cite{Seiberg1999}. The gauge theories on NC
spaces are equivalent (the Seiberg-Witten map) to effective
commutative theories with the additional deformation parameters
$\theta_{\mu\nu}$. This allows us to formulate a Lagrange field
theory on NC spaces with the same degrees of freedom in terms of
ordinary fields. The effective Lagrangian is formulated in terms
of ordinary fields and the parameters $\theta_{\mu\nu}$. The
coordinates of NC spaces obey the commutation relation
\cite{Snyder}, \cite{Connes}:
$\left[\widehat{x}_\mu,\widehat{x}_\nu \right]=i\theta_{\mu\nu}$.
We use here the Lorentz-Heaviside units, and set $\hbar=c=1$.

The effective Lagrangian density in terms of ordinary fields,
which is equivalent to Maxwell's Lagrangian density on NC spaces,
in the leading order of $\theta$ is \cite{Bichl}
\begin{equation}
 {\cal L}=-\frac14F^2_{\mu \nu}+\frac18\theta_{\alpha\beta}
 F_{\alpha\beta}F^2_{\mu \nu}-\frac12\theta_{\alpha\beta}
 F_{\mu \alpha}F_{\nu \beta}F_{\mu \nu}+{\cal O}(\theta^2) .
\label{1}
\end{equation}
The field strength tensor is given by
\begin{equation}
 F_{\mu \nu }=\partial _\mu A_\nu -\partial _\nu A_\mu ,
\label{2}
\end{equation}
where $A_\mu =({\bf A},iA_0)$ is the electromagnetic field
vector-potential; $E_i =iF_{i4}$ is the electric field; $B_{i}
=\epsilon_{ijk}F_{jk}$ ($\epsilon_{123}=1$) is the magnetic
induction field.

The BRST-shift symmetry and the vacuum polarization of photons at
the one-loop level of the theory under consideration have been
investigated in \cite{Fruhwirth}. Authors of \cite{Jackiw},
\cite{Guralnic} showed that the velocity of photon propagation in
the direction that is perpendicular to a background magnetic
induction field in such a $\theta$-deformed Maxwell theory differs
from $c$. The Lagrangian (1) can also be rewritten as follows:
\begin{equation}
 {\cal L}=\frac12 \left( {\bf E}^2-{\bf B}^2 \right)\left[1+({\bf
 \theta}\cdot{\bf B})\right]-\left({\bf \theta}\cdot{\bf E}\right)\left({\bf
 E}\cdot{\bf B}\right)
 +{\cal O}(\theta^2) ,
\label{3}
\end{equation}
where $\theta_i=(1/2)\epsilon_{ijk}\theta_{jk}$, and we put
$\theta_{i4}=0$ which guarantees  the unitarity \cite{Gomis},
\cite{Aharony}. Parameters $\theta_{\mu\nu}$ are the CP violating
variables (at CP transformations
$\theta_{\mu\nu}\rightarrow-\theta_{\mu\nu}$), and, as a result,
particles possess the dipole moments \cite{Riad2000},
\cite{Sheikh2000}. Quantum electrodynamics on NC spaces is a
renormalizable theory \cite{Hayakawa2000}, \cite{Bichl} with the
conservation of CPT symmetry.

From Eq. (1) equations of motion \cite{Jackiw}, \cite{Guralnic}
read
\begin{equation}
\frac{\partial}{\partial t}{\bf D}-\mbox{rot} {\bf H}=0
,\hspace{0.3in} \mbox{div}{\bf D}=0 , \label{4}
\end{equation}
where $(\mbox{rot} {\bf H})_{i} =\epsilon_{ijk}\partial_{j}H_{k}$
and $\mbox{div}{\bf D}=\partial_{i}D_{i}$. The displacement (${\bf
D}$) and magnetic (${\bf H}$) fields are given by
\[
 {\bf D}={\bf E}+{\bf d},\hspace{0.3in}{\bf d}=({\bf \theta}\cdot
 {\bf B}){\bf E}-({\bf \theta}\cdot {\bf E}){\bf B}-
 ({\bf E}\cdot {\bf B}){\bf \theta} ,
\]
\vspace{-8mm}
\begin{equation}
\label{5}
\end{equation}
\vspace{-8mm}
\[
 {\bf H}={\bf B}+{\bf h},\hspace{0.3in}{\bf h}=({\bf \theta}\cdot
 {\bf B}){\bf B}+({\bf \theta}\cdot {\bf E}){\bf E}-
 \frac{1}{2}\left({\bf E}^2- {\bf B}^2\right){\bf \theta} .
\]
Eq. (2) leads to $ \partial_\mu \widetilde{F}_{\mu \nu }=0$ (where
$\widetilde{F}_{\mu\nu}=(1/2)\varepsilon _{\mu \nu \alpha \beta }F
_{\alpha \beta}$, $\varepsilon _{\mu \nu \alpha \beta }$ is an
antisymmetric tensor, $\varepsilon _{1234}=-i$) which is rewritten
as
\begin{equation}
\frac {\partial}{\partial t}{\bf B}+\mbox{rot}{\bf
E}=0,\hspace{0.3in} \mbox{div}{\bf B}=0 . \label{6}
\end{equation}
\section{ Energy-momentum Tensor}

Using the general procedure, we obtain the canonical conservative
energy-momentum tensor \cite{Kruglov}:
\[
T_{\mu\nu}=-F_{\mu\alpha}F_{\nu\alpha}\left(1-\frac12\theta_{\gamma\beta}
 F_{\gamma\beta}\right)+\frac14
 \theta_{\mu\alpha}F_{\nu\alpha}F^2_{\rho\beta}
\]
\vspace{-8mm}
\begin{equation}
\label{7}
\end{equation}
\vspace{-8mm}
\[
-\theta_{\mu\beta}F_{\gamma\nu} F_{\rho\beta}F_{\gamma\rho}-\left(
F_{\mu\alpha}F_{\nu\gamma}+F_{\nu\alpha}F_{\mu\gamma}\right)
\theta_{\alpha\beta}F_{\gamma\beta}-\delta_{\mu\nu}{\cal L} .
\]
The canonical energy-momentum tensor (7) is non-symmetric, but for
classical electrodynamics at $\theta\rightarrow 0$, it becomes
symmetric. The trace of the energy-momentum tensor (7) is not
equal to zero; that indicates the trace anomaly at the classical
level. This is the consequence of the violation of the dilatation
symmetry: $x'_\mu=\lambda x_\mu$ for NC spaces.

Using the general procedure \cite{Landau}, and varying the action
corresponding to the Lagrangian (1) on the metric tensor, we find
the symmetric energy-momentum tensor:
\begin{equation}
T^{sym}_{\mu\nu}=T_{\mu\nu}+\frac14 \theta_{\nu\alpha}
 F_{\mu\alpha}F^2_{\rho\beta}
-\theta_{\nu\beta}F_{\gamma\mu} F_{\rho\beta}F_{\gamma\rho} .
\label{8}
\end{equation}
Here the conservative tensor $T_{\mu\nu}$ is defined by Eq. (7).
The components of the stress tensor tensor, found from Eq. (8),
are given by
\begin{equation}
T^{sym}_{44}=\frac {{\bf E}^2 +{\bf B}^2}{2}\left[1+({\bf
\theta}\cdot {\bf B})\right]-\left({\bf E}\cdot{\bf B}\right)
\left({\bf \theta}\cdot{\bf E}\right) . \label{9}
\end{equation}
\begin{equation}
T^{sym}_{m4}=-i\left\{\left[1+({\bf \theta}\cdot{\bf
B})\right]({\bf E}\times{\bf B})+\frac12 \left({\bf B}^2-{\bf
E}^2\right)({\bf E}\times {\bf \theta})\right\}_m .\label{10}
\end{equation}
\[
T^{sym}_{mn}=E_m E_n +B_m B_n -\frac12 \delta_{mn}\left( {\bf
E}^2+{\bf B}^2\right) +\left( {\bf \theta}\cdot{\bf
B}\right)\left(3E_m E_n +B_m B_n\right)
\]
\[
+\frac12\left( {\bf E}^2+{\bf B}^2\right)\left[B_m
\theta_n+\theta_m B_n - 3\delta_{mn}\left( {\bf \theta}\cdot{\bf
B}\right)\right]
\]
\vspace{-8mm}
\begin{equation}
\label{11}
\end{equation}
\vspace{-8mm}
\[
-\left( {\bf E}\cdot{\bf B}\right)\left[\theta_m E_n+\theta_n E_m
- \delta_{mn}\left( {\bf \theta}\cdot{\bf E}\right)\right]-\left(
{\bf \theta}\cdot{\bf E}\right)\left(E_m B_n+E_n B_m\right)
 \]
\[
-\left({\bf E}\times{\bf \theta}\right)_m\left({\bf B}\times{\bf
E}\right)_n-\left({\bf E}\times{\bf \theta}\right)_n\left({\bf
B}\times{\bf E}\right)_m .
\]
Eqs. (9)-(11) lead to the trace (anomaly) of the symmetric
energy-momentum tensor:
\begin{equation}
T^{sym}_{\mu\mu}=2\left( {\bf \theta}\cdot{\bf B}\right)\left({\bf
E}^2-{\bf B}^2\right)-4\left( {\bf \theta}\cdot{\bf
E}\right)\left({\bf E}\cdot{\bf B}\right) .
 \label{12}
\end{equation}
From Eqs. (7), (12), we find the relation
$T^{sym}_{\mu\mu}=2T_{\mu\mu}$. The trace anomaly can contribute
to the cosmological constant, and as a result, this trace anomaly
should be taken into account by consideration of the inflation
theory.

The trace anomaly vanishes in the case of the plane
electromagnetic waves as two Lorentz invariants ${\bf E}^2-{\bf
B}^2$, $({\bf E}\cdot{\bf B})$ are equal to zero, as well as for
classical electrodynamics at $\theta=0$.
\section{ Dirac's Quantization}

In accordance with the general procedure \cite{Dirac} of gauge
covariant quantization, we find from Eq. (1) (in leading order of
$\theta)$) the momenta:
\[
 \pi_i=\frac{\partial {\cal L}}{\partial(\partial_0 A_i)}=-E_i\left[1+({\bf
 \theta}\cdot{\bf B})\right]+\left({\bf \theta}\cdot{\bf E}\right)
  B_i+\left({\bf E}\cdot{\bf B}\right)\theta_i  ,
\]
\vspace{-8mm}
\begin{equation}
\label{13}
\end{equation}
\vspace{-8mm}
\[
\pi_0=\frac{\partial {\cal L}}{\partial (\partial_0 A_0)}=0 .
\]
A primary constraint follows from Eq. (13):
\begin{equation}
 \varphi_1 (x)=\pi_0 ,\hspace{0.3in}\varphi_1 (x)\approx 0 .
\label{14}
\end{equation}
Here Dirac's symbol $\approx$ is used for equations which hold
only weakly. Taking into account the equality $\pi_i=-D_i$, we
find the Poisson brackets
\begin{equation}
 \{A_i (\textbf{x},t),D_j(\textbf{y},t)\}=-\delta_{ij}
 \delta(\textbf{x}-\textbf{y}) ,
\label{15}
\end{equation}
\begin{equation}
 \{B_i (\textbf{x},t),D_j(\textbf{y},t)\}=\epsilon_{ijk}\partial_k
 \delta(\textbf{x}-\textbf{y}) .
\label{16}
\end{equation}
In the quantum theory, we should make the replacement
$\{B,D\}\rightarrow -i\left[B,D\right]$, where
$\left[B,D\right]=BD-DB$ is the commutator. Using Eqs. (3), (13),
and the relation ${\cal H}=\pi_\mu\partial_0 A_\mu-{\cal L}$, we
arrive at the density of the Hamiltonian:
\begin{equation}
 {\cal H}=\frac12 \left( {\bf E}^2+{\bf B}^2 \right)\left[1+({\bf
 \theta}\cdot{\bf B})\right]-\left({\bf \theta}\cdot{\bf E}\right)\left({\bf
 E}\cdot{\bf B}\right)-\pi_m \partial_m A_0 .
\label{17}
\end{equation}
As the primary constraint (14) have to be a constant of motion, we
find the condition
\begin{equation}
 \partial_0 \pi_0 =\{\pi_0,H\}=-\partial_m \pi_m= 0 ,
\label{18}
\end{equation}
where $H=\int d^3x {\cal H}$ is the Hamiltonian. Eq. (18) leads to
the secondary constraint:
\begin{equation}
 \varphi_2 (x)=\partial_m \pi_m ,\hspace{0.3in}\varphi_2 (x)\approx 0
 .
\label{19}
\end{equation}
This is simply the Gauss law (see Eq. (4)). It is easy to find $
\partial_0 \varphi_2 =\{\varphi_2,H\}\equiv 0$ , i.e. no additional
constraints. There are no second class constrains here because
$\{\varphi_1,\varphi_2\}=0$, as similar to classical
electrodynamics \cite{Dirac}. In accordance with the general
approach \cite{Dirac}, to have the total density of Hamiltonian,
we add to Eq. (17) the Lagrange multiplier terms:
\begin{equation}
 {\cal H}_T={\cal H}+v(x)\pi_0+u(x)\partial_m \pi_m .
\label{20}
\end{equation}
The density energy, obtained from Eq. (17), is ${\cal E}\equiv
T^{sym}_{44}$ (see Eq. (9)). The electric field, with the accuracy
of ${\cal O}(\theta^2)$, is given by
\begin{equation}
E_i=-\pi_i\left[1-({\bf
 \theta}\cdot{\bf B})\right]-\left({\bf \theta}\cdot{\bf \pi}\right)
  B_i-\left({\bf \pi}\cdot{\bf B}\right)\theta_i  ,
\label{21}
\end{equation}
so that the total density of Hamiltonian becomes
\[
 {\cal H}_T=\frac {{\bf \pi}^2+{\bf B}^2}{2}+({\bf
 \theta}\cdot{\bf B})\frac {{\bf B}^2-{\bf \pi}^2}{2}
 +\left({\bf \theta}\cdot{\bf \pi}\right)\left({\bf
 \pi}\cdot{\bf B}\right)
\]
\vspace{-8mm}
\begin{equation}
\label{22}
\end{equation}
\vspace{-8mm}
\[
 +v(x)\pi_0+\left(u(x)+A_0\right)
 \partial_m \pi_m ,
\]
where ${\bf B}=\mbox{rot}{\bf A}$.

Using Eq. (22), we obtain the Hamiltonian equations
\[
\partial_0 A_i=\{A_i,H\}=\frac{\delta H}{\delta \pi_i}
\]
\vspace{-8mm}
\begin{equation}
\label{23}
\end{equation}
\vspace{-8mm}
\[
=\pi_i\left[1-({\bf \theta}\cdot{\bf B})\right]+\left({\bf
B}\cdot{\bf \pi}\right) \theta_i+\left({\bf \pi}\cdot{\bf
\theta}\right)B_i -\partial_i A_0-\partial_i u(x) ,
\]
\[
\partial_0 \pi_i=\{\pi_i, H\}=-\frac{\delta H}{\delta A_i}
=\partial_n\left\{\left[ (\partial_n A_i)-(\partial_i A_n)\right]
\left[1+({\bf
 \theta}\cdot{\bf B})\right]\right\}
\]
\vspace{-8mm}
\begin{equation}
\label{24}
\end{equation}
\vspace{-8mm}
\[
 +\varepsilon_{iab}\partial_b\left[
\theta_a\frac {{\bf B}^2-{\bf \pi}^2}{2}+\pi_a ({\bf \pi}\cdot{\bf
\theta})\right] ,
\]
\begin{equation}
\partial_0 A_0=\{A_0,H\}=\frac{\delta H}{\delta \pi_0}=v(x),\hspace{0.1in}
\partial_0 \pi_0=\{\pi_0,H\}=-\frac{\delta H}{\delta
A_0}=-\partial_m \pi_m , \label{25}
\end{equation}
which are gauge-equivalent to the Euler-Lagrange equations. Taking
into account the definition (5), we find that Eq. (24) coincides
with the first equation in (4), and Eq. (23) is the gauge
covariant form of Eq. (21). Gauss's law is the secondary
constraint in this Hamiltonian formalism and is the second
equation in (4). The first class constraints generate gauge
transformations. In accordance with Eq. (25), the $A_0$ is
arbitrary function and Eqs. (25) represent the time evolution of
non-physical fields. The variables $\pi_0$, $\partial_m \pi_m$
equal zero as constraints.
\section{ The Coulomb Gauge}

Using the gauge freedom here the radiation gauge will be
considered. It should be noted that the gauge fixing approach is
beyond the Dirac's approach. Using the gauge freedom (described by
functions
 $v(x)$, $u(x)$), we impose new constraints:
\begin{equation}
\varphi_3 (x)=A_0\approx 0 ,\hspace{0.3in}\varphi_4 (x)=\partial_m
A_m\approx 0 . \label{26}
\end{equation}
Then equations (see also \cite{Hanson})
\begin{equation}
\{\varphi_1 (\textbf{x},t),\varphi_3
(\textbf{y},t)\}=-\delta(\textbf{x}-\textbf{y}) ,\hspace{0.1in}
\{\varphi_2 (\textbf{x},t),\varphi_4
(\textbf{y},t)\}=\Delta_x\delta(\textbf{x}-\textbf{y}) ,
\label{27}
\end{equation}
hold, where $\Delta_x\equiv\partial^2/(\partial x_m)^2$. Defining
``coordinates" $Q_i$ and conjugated momenta $P_i$ ($i=1,2$):
\begin{equation}
Q_i=(A_0,\partial_m
A_m),\hspace{0.3in}P_i=(\pi_0,-\Delta^{-1}_x\partial_m \pi_m) ,
\label{28}
\end{equation}
we find
\begin{equation}
\{Q_i(\textbf{x},t),P_j(\textbf{y},t)\}=\delta_{ij}
\delta(\textbf{x}-\textbf{y}) ,\hspace{0.3in}
\Delta^{-1}_x=-\frac{1}{4\pi|\textbf{x}|} ,\hspace{0.3in}\Delta_x
\frac{1}{4\pi|\textbf{x}|}=-\delta(\textbf{x}) ,\label{29}
\end{equation}
Canonical variables $Q_i,P_i$, Eq. (29) are not the physical
degrees of freedom and must be eliminated. Using the matrix of
Poisson brackets \cite{Hanson} $C_{ij}=\{\varphi_i
(\textbf{x},t),\varphi_j (\textbf{y},t)\}$, and the definition of
the Dirac brackets \cite{Hanson}, \cite{Henneaux}, we obtain
\begin{equation}
\{\pi_0(\textbf{x},t),A_0(\textbf{y},t)\}^*=
\{\pi_0(\textbf{x},t),A_i(\textbf{y},t)\}^*=
\{\pi_i(\textbf{x},t),A_0(\textbf{y},t)\}^*=0 ,\label{30}
\end{equation}
\begin{equation}
\{\pi_i(\textbf{x},t),A_j(\textbf{y},t)\}^*=
-\delta_{ij}\delta(\textbf{x}-\textbf{y})+\frac{\partial^2}{\partial
x_i \partial y_j} \frac{1}{4\pi|\textbf{x}-\textbf{y}|}
\hspace{0.2in}(i,j=1,2,3) ,\label{31}
\end{equation}
\begin{equation}
\{\pi_\mu(\textbf{x},t),\pi_\nu(\textbf{y},t)\}^*=
\{A_\mu(\textbf{x},t),A_\nu(\textbf{y},t)\}^*=0
\hspace{0.3in}(\mu, \nu=1,2,3,4) .\label{32}
\end{equation}
Taking into consideration the Fourier transformation of the
Coulomb potential, Eq. (31) takes the form
\begin{equation}
\{\pi_i(\textbf{k}),A_j(\textbf{q})\}^*=-(2\pi)^3
\delta(\textbf{k+q})\left(\delta_{ij}-\frac{k_i
k_j}{\textbf{k}^2}\right)
 .\label{33}
\end{equation}
The projection operator in the right side of Eq. (33) extracts the
physical transverse components of vectors.

Following the general method, we can set all second class
constraints strongly to zero, and as a result, only transverse
components of the vector potential $A_\mu$ and momentum $\pi_\mu$
are physical independent variables. Pairs of operators (28) are
absent in the reduced physical phase space and the physical
Hamiltonian becomes $H^{ph}=\int d^3 x {\cal E}$. Equations of
motion obtained from this Hamiltonian coincide with Eqs. (23)-(25)
at $u(x)=v(x)=0$ (the Poisson brackets are replaced by the Dirac
brackets). In the quantum theory, we should substitute Dirac's
brackets by the quantum commutators according to the prescription
$\{.,.\}^*\rightarrow-i[.,.]$. The fields $\textbf{E}$,
$\textbf{B}$, $\textbf{D}$, $\textbf{H}$ are invariants of the
gauge transformations and are measurable quantities (observables).
\section{ Conclusion}

We have just considered the effective non-linear electrodynamics,
which is equivalent to Maxwell's theory on NC spaces due to the
Seiberg-Witten map. The canonical conservative, symmetric
non-conservative energy-momentum tensors, and their non-zero
traces were found. We have shown that there is a trace anomaly
which is related with the violation of the dilation symmetry at
the classical level. The trace anomaly is absent in the case of
the plane electromagnetic waves. At high energy (on short
distances), when space-time might be non-commutative, the trace
anomaly can contribute to the cosmological constant, and as a
result, effect cosmology and physics of early universe.

Dirac's quantization of the effective non-linear electromagnetic
theory is similar to the quantization of classical
electrodynamics, and includes first class constraints. The gauge
fixing approach, on the basis of Coulomb's gauge, leads to second
class constraints and transition from the Poisson brackets to the
Dirac brackets. For consideration of the vacuum state one needs to
construct the Fock representation or the wave functionals. The
normalization conditions are formulated here in terms of
functional integrals with the corresponding gauge conditions. This
may involve, however, the introducing ghosts and negative norm
states, which is beyond the Dirac approach. The gauge covariant
Dirac quantization does not violate the Lorentz invariance and
locality in space, and has an advantage compared to the reduced
phase space approach (see \cite{Henneaux}).

The quantization of the Maxwell theory on NC spaces within
BRST-scheme was investigated in \cite{Bichl}, \cite{Fruhwirth},
\cite{Wulkenhaar}.

\end{document}